# Semi-Quantum Inspired Lightweight Mediated Quantum Key Distribution with Limited Resource and Untrusted TP

Cheng-Ching Kuo[1] and Tzonelih Hwang[2] (corresponding author)

[1,2]*Department of Computer Science and Information Engineering, National Cheng Kung University, No. 1, University Rd., Tainan City, 70101, Taiwan, R.O.C.*

[1] p76071072@mail.ncku.edu.tw

[2] hwangtl@csie.ncku.edu.tw

**Abstract**—Semi-quantum inspired lightweight protocol is an important research issue in realization of quantum protocols. However, the previous semi-quantum inspired lightweight mediated quantum key distribution (SQIL-MQKD) protocols need to use the Bell states or measure the Bell states. The generation and measurement of Bell states are more difficult and expensive than those of single photons. To solve this problem, a semi-quantum inspired lightweight mediated quantum key distribution with limited resource protocol is proposed. In this protocol, an untrusted third party (TP) only needs to perform the quantum operations related to single photons and the participants only have to perform two quantum operations: (1) reflecting qubits without disturbance (2) performing unitary operations on single photons. In addition, this protocol is showed to be robust under the collective attack.

**Keywords**: Quantum Cryptography; Quantum key distribution; Semi-quantum inspired lightweight quantum protocol; Untrusted third-party; Circular transmission

## 1. Introduction

In 2009, Boyer et al. [1] proposed the first semi-quantum key distribution (SQKD)

protocol which is an important research of quantum cryptography. The main goal of SQKD protocol is to let a quantum participant share a secret key with a classical participant who is limited to perform the following restricted operations: (1) generating qubits in Z-basis($|0\rangle, |1\rangle$), (2) measuring qubits with Z-basis, (3) reordering the qubits and (4) reflecting qubits without disturbance. Subsequently, various kinds of SQKD protocols [2-5] have been proposed.

In 2015, Krawec et al. [6] proposed the first mediated SQKD protocol, which allows two classical participants to share a secret key with the help of an untrusted (dishonest) third party (TP). Later, Hwang et al. [7] proposed the first semi-quantum inspired lightweight mediated quantum key distribution (SQIL-MQKD) protocol which is inspired by the "semi-quantum environment". To reduce the quantum overhead of TP in Hwang et al.'s protocol, Kuo et al. [8] also proposed an SQIL-MQKD protocol using single photons as quantum resource. However, the quantum overhead of TP in this protocol is still heavy because TP needs to perform Bell measurement on two particles from two participants, which implies that the TP must have the delay lines to store a photon until the reception of the other participant's particle.

To further reduce TP's burden, a new SQIL-MQKD protocol is proposed here, in which TP only needs to generate the qubits in X basis ($|+\rangle$ or $|-\rangle$) and measure the qubits in X basis (i.e., TP does not need a delay line) and the participants only have two quantum capabilities: (1) reflecting qubits without disturbance (2) performing unitary operations on single photons. This contribution makes our protocol easier to implement since all participants in the protocol could be the TP (quantum sever).

The rest of this paper is organized as follows. Section 2 describes the relationship between the single photon and unitary operations. Section 3 presents the limited resource semi-quantum inspired lightweight mediated quantum key distribution protocol. Section 4 shows the proposed protocol is robust. In Section 5, a comparison is given with the existing three-party QKD protocols. Finally, a conclusion is given in Section 6.

## 2. Unitary operators and Quantum state transition

Before describing the proposed protocol, we first review some background which will be useful to understand the protocol. Unitary operators, important in quantum computation and quantum information, can perform on a single photon and transform the state of a photon to the other state.

This section introduces well-known unitary operations and the state transitions of single photons by them as follows. In addition, we will use the relationship shown in Table 1 to design the protocol in the next section.

Identity operator ($I$):

The identity matrix leaves any state unchanged (i.e., do nothing). The corresponding matrix representation: $I = |0\rangle\langle 0| + |1\rangle\langle 1| = \begin{bmatrix} 1 & 0 \\ 0 & 1 \end{bmatrix}$.

Pauli-Z operator ($\sigma_z$):

The Z Pauli matrix leaves $|0\rangle$ unchanged, whereas takes $|1\rangle$ to $-|1\rangle$. As for an X-basis state, ($|+\rangle$ or $|-\rangle$), it converts from one (e.g. $|+\rangle$) to the other ($|-\rangle$). The corresponding matrix representation: $\sigma_z = |0\rangle\langle 0| - |1\rangle\langle 1| = \begin{bmatrix} 1 & 0 \\ 0 & -1 \end{bmatrix}$.

Hadamard operator ($H$):

$H$ takes a Z-basis state ($|0\rangle$ or $|1\rangle$) to an X-basis state ($|+\rangle$ or $|-\rangle$) and an X- basis state into a Z-basis state. It is represented by the Hadamard matrix: $H = \frac{1}{\sqrt{2}}(|0\rangle\langle 0| + |0\rangle\langle 1| + |1\rangle\langle 0| - |1\rangle\langle 1|) = \frac{1}{\sqrt{2}}\begin{bmatrix} 1 & 1 \\ 1 & -1 \end{bmatrix}$.

**Table 1** The state transitions of single photons

| Unitary operation \ State | $\|0\rangle$ | $\|1\rangle$ | $\|+\rangle$ | $\|-\rangle$ |
|---|---|---|---|---|
| $I$ | $\|0\rangle$ | $\|1\rangle$ | $\|+\rangle$ | $\|-\rangle$ |
| $\sigma_z$ | $\|0\rangle$ | -$\|1\rangle$ | $\|-\rangle$ | $\|+\rangle$ |
| $H$ | $\|+\rangle$ | $\|-\rangle$ | $\|0\rangle$ | $\|1\rangle$ |

## 3. Proposed Limited Resource SQIL-MQKD Protocol

This section describes a limited resource SQIL-MQKD protocol allowing two lightweight quantum users to share a secret key with the help of an untrusted TP. The participants, Alice and Bob, only has two quantum capabilities: (1) reflecting the qubits without any interruption, and (2) performing unitary operations on a qubit. Besides, TP only needs to prepare the single photons and perform X-basis measurement.

There are ideal quantum channels (i.e., non-lossy and noiseless) between TP and Alice, TP and Bob, Alice and Bob. The classical channel connected between Alice and Bob is assumed to be authenticated (no one can modify the content of the published message). TP is still untrusted which implies that TP may perform any possible attacks to compromise the shared key. The steps of this protocol are described as follows (also shown in Figure 1):

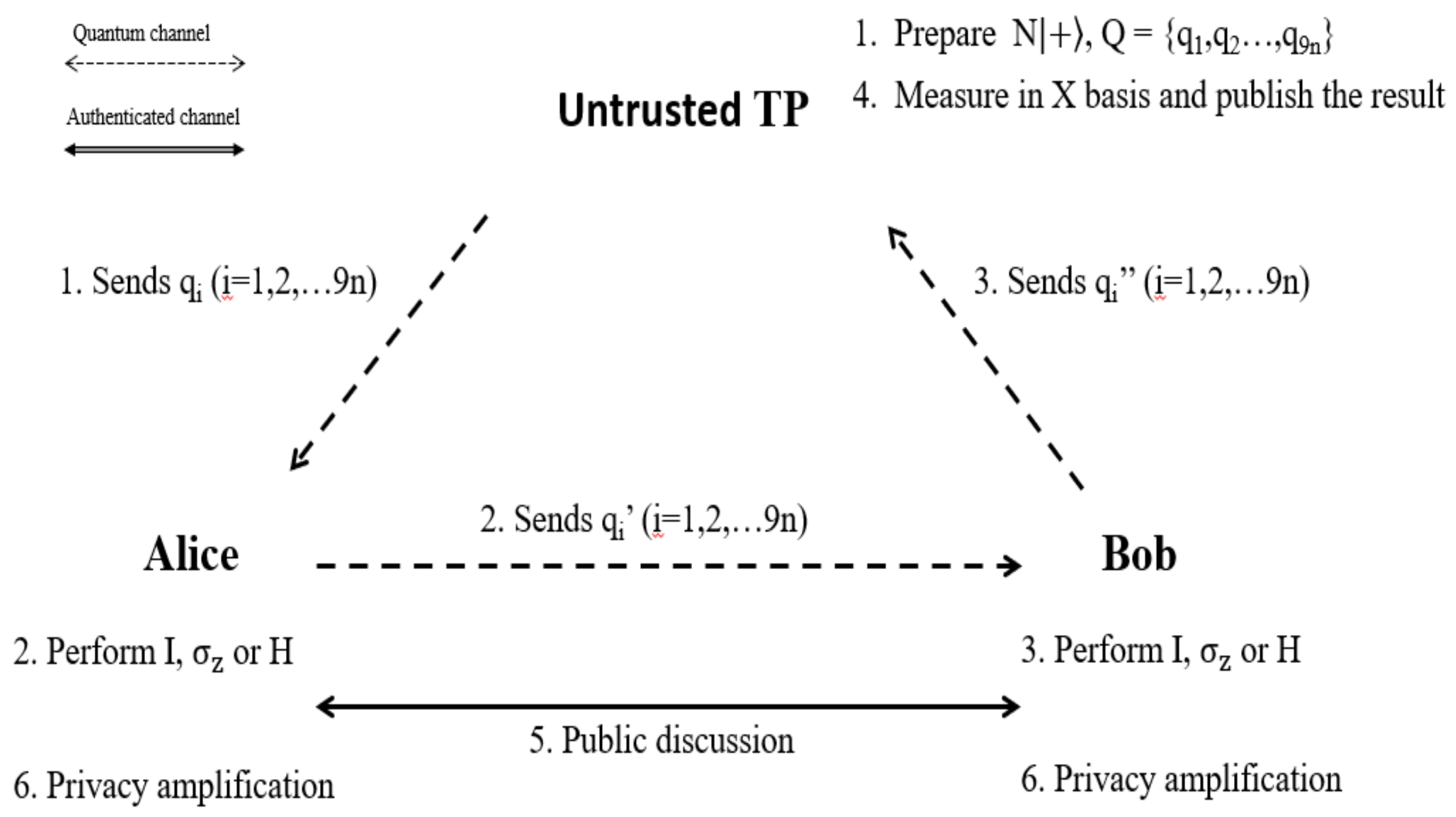

**Figure 1** Proposed Limited Resource SQIL-MQKD protocol

**Step 1** TP prepares $N(=9n)$ $|+\rangle$ ($Q=\{q_1, q_2, ..., q_{9n}\}$), and sends them to Alice one-by-one..

**Step 2** Upon receiving a qubit $q_i$ ($i=1,2,...,9n$) from TP, Alice chooses to perform one of the unitary operations ($I$, $\sigma_z$, $H$) on it and sends the result qubit, $q_i'$, to Bob.

**Step 3** When Bob receives a qubit $q_i'$ ($i=1,2,...,9n$) from Alice, Bob also chooses to perform one of the unitary operations ($I$, $\sigma_z$, $H$) on it and sends the result qubit, $q_i''$, back to TP.

**Step 4** Then TP measures $q_i''$ ($i=1,2,...,9n$) in X basis and records the measurement result. After all $q_i''$ are measured, TP publishes all of the measurement results.

**Step 5** Upon receiving the measurement results, Alice and Bob will start a discussion about the operations they had done in the previous step via an authenticated classical channel. Similar to the previous protocol, Alice will tell Bob whether she performs the Hadamard operation, and so will Bob. According to the information from Alice and Bob, three cases can be summarized as follows:

**Case 1.** When both Alice and Bob choose to perform the Hadamard operation, the qubit TP measures should be $H(H(|+\rangle))$ and they can check whether the measurement result is $|+\rangle$ or not. If the measurement result is not $|+\rangle$, it implies that there is an eavesdropping. If the error rate is higher than a pre-defined threshold value, Alice and Bob will abort the current protocol.

**Case 2.** Both Alice and Bob choose to perform $\sigma_z$ operation or $I$ operation (i.e., do nothing) before sending the qubit. In this case, Alice can get the key bit according to her operation ($I$ : 0, $\sigma_z$ : 1). As for Bob, he can get the same key bit as Alice's according to his operation and TP's measurement result. For example, suppose that Bob performed $I$ operation and TP's measurement result is $|+\rangle$. Then Bob knows Alice's operation is $I$ and key bit is 0. On the other hand, if Bob performed $I$ operation and TP's measurement result is $|-\rangle$, then he knows Alice's operation is $\sigma_z$ and key bit is 1. (as also shown in **Table 2**)

**Table 2** The sharing key bit

| **Alice's operation** | **Bob's operation** | **Expected measurement result** | **Shared key bit** |
|---|---|---|---|
| $I$ | $I$ | $\|+\rangle$ | 0 |
| $\sigma_z$ | $I$ | $\|-\rangle$ | 1 |
| $I$ | $\sigma_z$ | $\vert -\rangle$ | 0 |
| $\sigma_z$ | $\sigma_z$ | $\vert +\rangle$ | 1 |

**Case 3.** Except for the situations in Case 1 and Case 2, Alice and Bob will discard the measurement results because the qubit is measured in incorrect basis. In other word, Bob cannot use this case to derive Alice's operation.

**Step 6** After discussing all the results, Alice and Bob randomly choose half of the secret key bits and disclose the values to check the eavesdroppers and TP's faithfulness in executing the protocol. If the chosen values of bits are the same, the remaining bits are treated as the shared secret key after they perform privacy amplification [9-10]. If not, Alice and Bob will terminate the protocol and start from the beginning.

## 4. Security analyses

In this section, we will analyze the security of the proposed protocol. To be specific, we will show the robustness of the proposed protocol [1]. To prove the proposed protocol is robust, we have to prove that it is free from the collective attack, which is a very important class of attacks because it includes almost all the well-known attacks [11-12], such as measurement attack and intercept-resend attack, and etc.

### Collective attack

With the collective attack, the attacker will perform some unitary operators to entangle his ancillary qubits with the original quantum. Then, the attacker measures the ancillary qubits to obtain useful information. We assume that TP is the attacker (Eve) because TP has more authority than other eavesdroppers in the protocol. If the proposed protocol can resist the collective attack from a malicious TP, it can also resist the attack from other eavesdroppers. In this study, we want to prove that TP cannot obtain useful information without being detected. That is, if TP wants to obtain useful information, he/she will introduce a detectable interruption to the original system and hence the attack will be detected.

**Theorem 1:** TP, in the proposed protocol, cannot obtain any useful information by the collective attack.

**<Proof>:** We assume that TP generates the ancillary qubits $E=\{|E_1\rangle,|E_2\rangle,...\}$ and performs a unitary operation $U_1$ on each qubit in Step 1, where the operation $U_1$ must satisfy $U_1 * U_1 = I$. After Alice and Bob finishing the operations in Step 2 and Step 3, TP performs another unitary operations $U_2$ and $U_3$ on each qubit send from Alice and Bob, respectively. Then TP can measure the ancillary qubits to obtain the measurement results and deduce the values of the shared key bits with the information of the ancillary qubits. First, we define $U_1$, $U_2$ and $U_3$ operation as follow:

$$U_1|+\rangle_A \otimes |E_1\rangle_E = a_1|+\rangle|e_1\rangle + a_2|-\rangle|e_2\rangle \qquad \text{Eq. (1)}$$

where $|E_1\rangle$ denotes the initial state of ancillary qubit; $|a_1|^2+|a_2|^2=1$; $|e_1\rangle$ and $|e_2\rangle$ are two states which are distinguishable to TP.

$$\begin{aligned}
&U_2|+\rangle_A \otimes |E_2\rangle_E = A_1|+\rangle|F_1\rangle + A_2|-\rangle|F_2\rangle \\
&U_2|-\rangle_A \otimes |E_2\rangle_E = B_1|+\rangle|G_1\rangle + B_2|-\rangle|G_2\rangle \\
&U_2|0\rangle_A \otimes |E_2\rangle_E = U_2(\frac{1}{\sqrt{2}}(|+\rangle+|-\rangle)) \otimes |E_2\rangle_E = \frac{1}{\sqrt{2}}(U_2|+\rangle \otimes |E_2\rangle_E + U_2|-\rangle \otimes |E_2\rangle_E)
\end{aligned} \qquad \text{Eq. (2)}$$

where $|E_2\rangle$ denotes the initial state of ancillary qubit; $|F_1\rangle$, $|F_2\rangle$, $|G_1\rangle$, $|G_2\rangle$ are four states which are distinguishable to TP; $|A_1|^2+|A_2|^2=1$ and $|B_1|^2+|B_2|^2=1$.

$$\begin{aligned}
&U_3|+\rangle_B \otimes |E_3\rangle_E = C_1|+\rangle|H_1\rangle + C_2|-\rangle|H_2\rangle \\
&U_3|-\rangle_B \otimes |E_3\rangle_E = D_1|+\rangle|K_1\rangle + D_2|-\rangle|K_2\rangle
\end{aligned} \qquad \text{Eq. (3)}$$

where $|E_3\rangle$ denotes the initial state of ancillary qubit; $|H_1\rangle$, $|H_2\rangle$, $|K_1\rangle$, $|K_2\rangle$ are

four states which are distinguishable to TP; $|C_1|^2+|C_2|^2=1$ and $|D_1|^2+|D_2|^2=1$ .

After Alice's and Bob's operations, TP computes all possible equations by Eq. (1), Eq. (2) and Eq. (3). The equations are shown as below. Note that the equations in case 3 are discarded.

**Alice and Bob both perform $I$:**

$$\begin{aligned}&U_2(a_1|+\rangle+a_2|-\rangle)\otimes|E_2\rangle_E\\&=a_1(A_1|+\rangle|F_1\rangle+A_2|-\rangle|F_2\rangle)+a_2(B_1|+\rangle|G_1\rangle+B_2|-\rangle|G_2\rangle)\end{aligned}$$

$$\begin{aligned}&U_3(a_1A_1|+\rangle+a_1A_2|-\rangle+a_2B_1|+\rangle+a_2B_2|-\rangle)\otimes|E_3\rangle_E\\&=(a_1A_1+a_2B_1)(C_1|+\rangle|H_1\rangle+C_2|-\rangle|H_2\rangle)\\&+(a_1A_2+a_2B_2)(D_1|+\rangle|K_1\rangle+D_2|-\rangle|K_2\rangle)\end{aligned}\qquad \text{Eq. (4),}$$

**Alice performs $I$, then Bob performs $\sigma_z$:**

$$\begin{aligned}&U_2(a_1|+\rangle+a_2|-\rangle)\otimes|E_2\rangle_E\\&=a_1(A_1|+\rangle|F_1\rangle+A_2|-\rangle|F_2\rangle)+a_2(B_1|+\rangle|G_1\rangle+B_2|-\rangle|G_2\rangle)\end{aligned}$$

$$\begin{aligned}&U_3(a_1A_1|-\rangle+a_1A_2|+\rangle+a_2B_1|-\rangle+a_2B_2|+\rangle)\otimes|E_3\rangle_E\\&=(a_1A_2+a_2B_2)(C_1|+\rangle|H_1\rangle+C_2|-\rangle|H_2\rangle)\\&+(a_1A_1+a_2B_1)(D_1|+\rangle|K_1\rangle+D_2|-\rangle|K_2\rangle)\end{aligned}\qquad \text{Eq. (5),}$$

**Alice performs $\sigma_z$, then Bob performs $I$:**

$$\begin{aligned}&U_2(a_1|-\rangle+a_2|+\rangle)\otimes|E_2\rangle_E\\&=a_1(B_1|+\rangle|G_1\rangle+B_2|-\rangle|G_2\rangle)+a_2(A_1|+\rangle|F_1\rangle+A_2|-\rangle|F_2\rangle)\end{aligned}$$

$$\begin{aligned}&U_3(a_1B_1|+\rangle+a_1B_2|-\rangle)+a_2A_1|+\rangle+a_2A_2|-\rangle)\otimes|E_3\rangle_E\\&=(a_1B_1+a_2A_1)(C_1|+\rangle|H_1\rangle+C_2|-\rangle|H_2\rangle)\\&+(a_1B_2+a_2A_2)(D_1|+\rangle|K_1\rangle+D_2|-\rangle|K_2\rangle)\end{aligned}\qquad \text{Eq. (6),}$$

**Alice performs $\sigma_z$, then Bob performs $\sigma_z$:**

$$\begin{aligned}&U_2(a_1|-\rangle+a_2|+\rangle)\otimes|E_2\rangle_E\\&=a_1(B_1|+\rangle|G_1\rangle+B_2|-\rangle|G_2\rangle)+a_2(A_1|+\rangle|F_1\rangle+A_2|-\rangle|F_2\rangle)\end{aligned}$$

$$\begin{aligned}&U_3(a_1B_1|-\rangle+a_1B_2|+\rangle)+a_2A_1|-\rangle+a_2A_2|+\rangle)\otimes|E_3\rangle_E\\&=(a_1B_2+a_2A_2)(C_1|+\rangle|H_1\rangle+C_2|-\rangle|H_2\rangle)\\&+(a_1B_1+a_2A_1)(D_1|+\rangle|K_1\rangle+D_2|-\rangle|K_2\rangle)\end{aligned}\qquad \text{Eq. (7),}$$

Now, TP has all possible equations after $U_2$ and $U_3$ operation in key case (i.e., case 2), then TP needs to compute the equation in check case (i.e., case 1) to get the limitation of ancillary qubits.

**Alice performs $H$, then Bob performs $H$:**

$$U_2(a_1|0\rangle + a_2|1\rangle)\otimes|E_2\rangle_E$$
$$=U_2(\frac{a_1+a_2}{\sqrt{2}}|+\rangle + \frac{a_1-a_2}{\sqrt{2}}|-\rangle)\otimes|E_2\rangle_E$$
$$=\frac{1}{\sqrt{2}}[(a_1+a_2)(\mathrm{A}_1|+\rangle|\mathrm{F}_1\rangle+\mathrm{A}_2|-\rangle|\mathrm{F}_2\rangle)+(a_1-a_2)(\mathrm{B}_1|+\rangle|\mathrm{G}_1\rangle+B_2|-\rangle|\mathrm{G}_2\rangle)]$$

$$U_3\{\frac{1}{\sqrt{2}}[(a_1+a_2)(\mathrm{A}_1|+\rangle|\mathrm{F}_1\rangle+\mathrm{A}_2|-\rangle|\mathrm{F}_2\rangle)+(a_1-a_2)(\mathrm{B}_1|+\rangle|\mathrm{G}_1\rangle+B_2|-\rangle|\mathrm{G}_2\rangle)]\}\otimes|E_3\rangle_E$$
$$=|+\rangle\otimes(\frac{a_1A_1+a_2A_1+a_1B_1-a_2B_1+a_1A_2+a_2A_2+a_1B_2-a_2B_2}{2}C_1|H_1\rangle+$$
$$\frac{a_1A_1+a_2A_1+a_1B_1-a_2B_1+a_1A_2+a_2A_2+a_1B_2-a_2B_2}{2}D_1|K_1\rangle) \qquad \text{Eq. (8)}$$
$$+|-\rangle\otimes(\frac{a_1A_1+a_2A_1+a_1B_1-a_2B_1+a_1A_2+a_2A_2+a_1B_2-a_2B_2}{2}C_2|H_2\rangle+$$
$$\frac{a_1A_1+a_2A_1+a_1B_1-a_2B_1+a_1A_2+a_2A_2+a_1B_2-a_2B_2}{2}D_2|K_2\rangle)$$

From Eq. (2), TP has to set $a_2|-\rangle|\mathrm{e}_2\rangle = A_2|-\rangle|F_2\rangle = B_1|-\rangle|G_2\rangle = C_2|-\rangle|\mathrm{H}_2\rangle = D_1|+\rangle|\mathrm{K}_1\rangle = \vec{0}$ to pass the Alice and Bob's detection. Similarly, from Eq. (8), if TP wants to pass the detection, he must set incorrect items to be $\vec{0}$. That is, TP has to set

$$(\frac{a_1A_1+a_2A_1+a_1B_1-a_2B_1+a_1A_2+a_2A_2+a_1B_2-a_2B_2}{2}C_2|H_2\rangle+$$
$$\frac{a_1A_1+a_2A_1+a_1B_1-a_2B_1+a_1A_2+a_2A_2+a_1B_2-a_2B_2}{2}D_2|K_2\rangle)=\vec{0}$$

Then, TP substitutes the limitation from Eq. (2) and get the simultaneous equations:

$(a_1A_1-a_1B_2)C_2|H_2\rangle+(a_1A_1-a_1B_2)D_2|K_2\rangle=0$ which can be deduced that $A_1=B_2$. Next, TP would substitute the deduced answer in Eq. (4-7) and get the following equations:

$$a_1A_1(C_1|+\rangle|H_1\rangle+C_2|-\rangle|H_2\rangle) \qquad Eq.\ (8),$$

$$a_1A_1(D_1\,|+\rangle\,|K_1\rangle + D_2\,|-\rangle\,|K_2\rangle) \qquad Eq.\ (9),$$

$$a_1B_2(\mathrm{D}_1\,|+\rangle\,|K_1\rangle + D_2\,|-\rangle\,|K_2\rangle) \qquad Eq.\ (10),$$

$$a_1B_2(C_1\,|+\rangle\,|H_1\rangle + C_2\,|-\rangle\,|H_2\rangle) \qquad Eq.\ (11).$$

Because TP can distinguish the states $|H\rangle, |K\rangle$, he/she would obtain some information about the participants' secret key bits from Eq. (8-11). That is, TP can know whether Alice's and Bob's operations are the same or not. However, in our protocol, this information is not useful for getting the key bits. In other words, TP cannot obtain any useful information by this attack. Based on these analyses, the robustness of our protocol is proven. That is, TP will introduce a detectable disturbance if he/she performs collective attack on the protocol.

## 5. Comparison

We compared the proposed protocol with the other mediated QKD protocols including Hwang et al.'s [7] and Yang et al.'s [13]. **Table 3** shows the comparison results.

In terms of TP's quantum capabilities, the TP in Hwang et al.'s protocol has to prepare Bell sates and perform Bell measurement. However, in the proposed protocol, TP only need to prepare single photons as the quantum resource and perform X-basis measurement. In terms of participant's quantum capabilities, our protocol allows both participants to perform only two simple quantum operations but Yang et al.'s protocol allows one of their participants to perform three quantum operations.

As for the qubit efficiency which is defined by the following equation: $Q = \frac{n}{m}$, n denotes the total number of shared key bits, and m denotes the number of total qubits used in the protocol. The proposed protocol is $\frac{2}{9}$. On the other hand, the qubit efficiencies of Hwang's scheme and Yang's scheme are $\frac{1}{9}$ and $\frac{1}{12}$, respectively.

**Table 3** Comparison with the other three-party QKD protocols

<table>
<tr><th></th><th>Hwang et al.'s [7]</th><th>Yang et al.'s [13]</th><th>Proposed protocol</th></tr>
<tr><td>TP's quantum capabilities</td><td>1.Perform Bell measurement<br>2.Prepare the Bell states</td><td>1.Perform X-basis measurement<br>2. Prepare the single photons in X basis</td><td>1.Perform X-basis measurement<br>2. Prepare the single photons in X basis</td></tr>
<tr><td rowspan="2">Participant's quantum capabilities</td><td rowspan="2">1.Unitary operation<br>2. Reflect</td><td>1.Prepare<br>2.Measure<br>3.Reflect</td><td rowspan="2">1.Unitary operation<br>2. Reflect</td></tr>
<tr><td>1.Unitary operation<br>2.Reflect</td></tr>
<tr><td>Qubit resource</td><td>Bell states</td><td>Single photons</td><td>Single photons</td></tr>
<tr><td>Qubit efficiency</td><td>$\frac{1}{9}$</td><td>$\frac{1}{12}$</td><td>$\frac{2}{9}$</td></tr>
</table>

## 6. Conclusions

This study proposes a limited resource semi-quantum inspired lightweight quantum key distribution protocol by using the properties of single photons. Because TP and participants only need to perform the quantum operations related to single photons, the proposed protocol is more practical than Hwang et al.'s protocol and Yang et al.'s protocol. In addition, based on the **Table 3**, the proposed protocol has the best qubit efficiency. And the security analysis shows that the proposed protocol allows two participants to share a secure key with the assistance of an untrusted TP. It is indeed an interesting future research idea to design other SQIL-MQKD protocols with various combinations of simple quantum capabilities.

## Acknowledgement

This research is partially supported by the Ministry of Science and Technology, Taiwan, R.O.C., under the Contract No. MOST 108-2221-E-006-107-; MOST 108-2627-E-006-001 -.

## References

[1] Boyer, M., Gelles, R., Kenigsberg, D., & Mor, T. (2009). Semiquantum key distribution. Physical Review A, 79(3).

[2] Zou, X., Qiu, D., Li, L., Wu, L., & Li, L. (2009). Semiquantum-key distribution using less than four quantum states. Physical Review A, 79(5).

[3] Boyer, M., & Mor, T. (2011). Comment on "Semiquantum-key distribution using less than four quantum states." Physical Review A, 83(4).

[4] Zou, X., Qiu, D., Zhang, S., & Mateus, P. (2015). Semiquantum key distribution without invoking the classical party's measurement capability. Quantum Information Processing, 14(8), 2981–2996.

[5] Yu, K.-F., Yang, C.-W., Liao, C.-H., Hwang, T.: Authenticated semi-quantum key distribution protocol using Bell states. Quantum Inf. Process. 13(6), 1457-1465 (2014).

[6] Krawec, W. O. (2015). Mediated semiquantum key distribution. Physical Review A, 91(3).

[7] Hwang, T., Chen, Y. J., Tsai, C. W., & Kuo, C. C. (2020). Semi-quantum Inspired Lightweight Mediated Quantum Key Distribution Protocol. arXiv preprint arXiv:2007.05804..

[8] Cheng-Ching Kuo, Tzonelih Hwang. Semi-Quantum Inspired Lightweight Mediated Quantum Key Distribution Using Single Photons.

[9] C. H. Bennett, G. Brassard, and J.-M. Robert, "Privacy amplification by public discussion," SIAM journal on Computing, vol. 17, no. 2, pp. 210-229, 1988.

[10] C. H. Bennett, G. Brassard, C. Crépeau, and U. M. Maurer, "Generalized privacy amplification," IEEE Transactions on Information Theory, vol. 41, no. 6, pp. 1915-1923, 1995.

[11] Biham, Boyer, Brassard, G. van de, and Mor, "Security of Quantum Key Distribution against All Collective Attacks," Algorithmica, vol. 34, no. 4, pp. 372-388, 2002/11/01 2002.

[12] V. Scarani, H. Bechmann-Pasquinucci, N. J. Cerf, M. Dušek, N. Lütkenhaus, and M. Peev, "The security of practical quantum key distribution," Reviews of Modern Physics, vol. 81, no. 3, pp. 1301-1350, 09/29/2009.

[13] Yang,Y,F Hwang, T. "Asymmetric semi-quantum security protocols: Semi-quantum key distribution and Semi-quantum secret sharing" master's thesis, NCKU CSIE ,2019. < https://hdl.handle.net/11296/3d948c>.